\documentclass{icrc}

\usepackage{times}
 \usepackage{graphicx} % when using Latex and dvips
\begin{document}

\title{Cracking Open the Window for Strongly Interacting Massive Particles as the Halo Dark Matter}
\author[1]{P.C. McGuire}
\affil[1]{Technical Faculty, University of Bielefeld, Bielefeld 33501, Germany }
\author[2]{P.J. Steinhardt}
\affil[2]{Department of Physics, Princeton University, Princeton, NJ 08544, USA}

\correspondence{mcguire@techfak.uni-bielefeld.de}

\firstpage{1}
\pubyear{2001}

% \titleheight{11cm} % uncomment and adjust in case your title block
		     % does not fit into the default and minimum 7.5 cm

\maketitle

\begin{abstract}
 In the early 1990's, an analysis was completed by several theorists of
the available mass/cross-section parameter space for unusual particle
candidates to solve the dark matter problem, e.g. strongly
interacting massive particles (SIMPs). This
analysis found several unconstrained windows, such that for SIMP masses
and cross-sections within these windows, SIMPs could still be the dominant
dark matter in our Galactic halo. Since the early 1990's, some of these
windows have been narrowed or closed, and some of these windows have been
widened further by more careful analysis. We summarize the present
state of the SIMP parameter space, and point to the cosmological salience
of SIMPs as dark matter, given some of the present inadequacies of the WIMP
solution to the dark matter problem.
\end{abstract}

\section{Cosmological Motivation for Trying to Jar the Window}
In the last 10-15 years considerable progress has been made
in the understanding of the cosmology of our universe. The original
1960's microwave measurements
of the residual background radiation (CMB) from the birth of our universe
were followed by precision measurements of this radiation
in the 1980's and 1990's. These new CMB measurements
point to a flat universe with ($\Omega_{\mathrm{TOT}}=1$).
Recent measurements of the absolute luminosities and redshifts of distant 
supernovae suggest a universe whose expansion is accelerating,
pointing to the existence of a repulsive dark energy,
accounting to about 70\% of the critical density.
And at least eight different measurement methods
suggest inpendently that the density of matter in the universe
(normal matter or otherwise) is more than a quarter {\it and}
less than half of the critical density (N. Bahcall {\it et al.}, 1999).

On a smaller distance (or redshift) scale than the cosmological
scale of the whole universe, we find that the normal baryonic
matter found in galaxies 
amounts to about 4\% of the critical density
 (only about 0.1\% in stars), where\-as
from the total-matter density-measurements mentioned
above, we believe that at least 25\% of the density
of the universe is composed of matter. This suggests
that there is a large amount of dark matter on a local galactic
scale to be identified and added to the equation at the 20-30\% level.

 \begin{figure*}[t]
 \includegraphics[width=17.0cm]{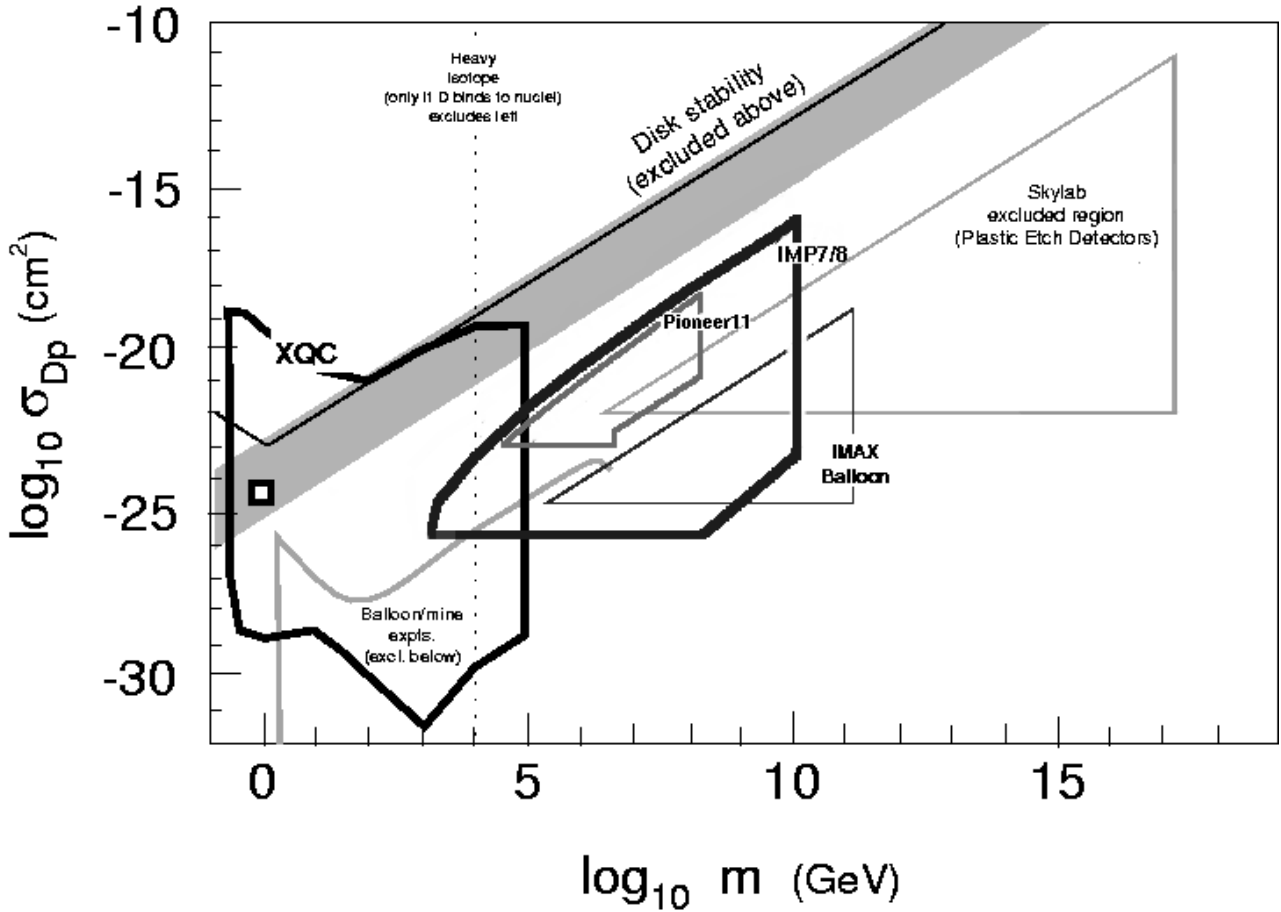}
  \caption{ This plot shows the constraints for dark matter-proton cross-section
 $\sigma_{Dp}$ versus mass obtained from a variety of experiments described
 in the text.  We assume here that dark matter (SIMPs) interactions
 with ordinary
 matter scale coherently with nucleon number of the scattering agent
 (the atmosphere or a detector).  The excluded regions are labeled as open
 or closed polygons.  The gray region  shows the range for the dark matter
 self-scattering ($\sigma_{DD}$ in Eq.~(1)) to avoid the small-scale structure
 problem with WIMPs.  For some SIMP candidates, $\sigma_{DD}$ and $\sigma_{Dp}$
may be comparable, although this is not required generally.     
}

 \end{figure*}

The standard solution to this for the past 20 years has
been Cold Dark Matter (CDM), in which the missing dark matter
is composed of subatomic collisionless particles (a.k.a.
weakly interacting massive particles (WIMPs)). WIMPs
can in principle be produced in abundance in the early universe 
so as to solve the dark matter problem.
WIMPs together with dark energy also seem to be
a good fit to the recent measurements of the spatial power spectrum
of microwave structure at 'large' angular scales of more than 0.2 degrees
on the sky.  

One possible problem of the WIMP solution
to the dark matter problem is that some simulations
suggest that WIMPs produce too much structure at
very small scales.  These simulations have
also shown that normal WIMP-like CDM produces 
dark galactic halos which have density profiles
with a divergence at the center of the halo.
These divergences have not been observed
in the real universe: galaxies tend to have
dark halo profiles that are flatter
than predicted. Also, these WIMP/CDM simulations
predict that there will be roughly 1000 satellite
galaxies in the Local Group of galaxies, where\-as
only about 100 are observed.

Spergel and Steinhardt (2000) have suggested one possible
solution: if the CDM has strong self-interactions (SIDM),
then the simulated over-abundance of small-scale
structure would not be a problem anymore. The strong
interactions would cause at least one scattering event
between SIDM particles during the history of the universe,
which would effectively smooth out the cuspy halos. They
even estimate an interaction cross-section
of colliding SIDM particles which
would alleviate the problems seen
in the simulations of WIMPs as CDM:
\begin{equation}
\begin{tabular}{lll}

$\frac{\sigma_{DD}}{m}$ & $ = $ &
 $8 \times 10^{-25}\, -\, 1 \times 10^{-23}\, \mathrm{cm}^2\mathrm{GeV}^{-1}$ \\
 & $=$ &
                   $0.5 \, - \, 6 \, \mathrm{cm}^2 \mathrm{g}^{-1}$,\\
\end{tabular}
\end{equation}
where $\sigma_{DD}$ is the elastic-scattering cross-section
be\-tween two SIDM part\-icles, and $m$ is the mass
of an SIDM particle. This region of SIDM parameter space is
shown as the gray band in Figure 1.

\section{The New Crack}
But SIDM may also have direct interactions of sim\-ilar strength
with normal matter. Starkman {\it et al.} (1990) analyzed
the observational and theoretical evidence for the
possibility that strongly-interacting massive
particles (SIMPs) 
could be the CDM in our galactic halo,
and yet remain undetected.
These SIMPs were assumed to have interactions with normal baryonic matter.
 The observational
and theoretical evidence analyzed by Starkman {\it et al.}
included a balloon-borne silicon detector
(Rich {\it et al.}, 1987), a silicon detector
on Pioneer11 (Simpson {\it et al.}, 1980),
plastic etch detectors on SKYLAB (Shirk and Price, 1978),
as well as an analysis of the stability of the Galactic
disk and an analysis of heavy water searches.

These SIMPs were studied by Starkman {\it et al.}
under two different scenarios, which had differing
interactions with normal baryonic matter.
One scenario had an interaction strength which couples to the number
of nucleons in a normal nucleus, and the other scenario
had an interaction strength which couples with the spin
of that nucleus. In both scenarios, they found
several open windows in cross-section parameter
space. This helped to spur on at least two SIMP search experiments
in the early 1990's (McGuire {\it et al.} (1994),
 Bacci {\it et al.} (1994)).

Meanwhile, McGuire (1994) noticed while re\-view\-ing Stark\-man {\it et al.}'s
work that several errors had been made, which opened
up a window for SIMPs. The main two errors were:
\begin{enumerate}
\item There was a  titanium shield of thickness 1.7 mg/cm$^2$
       in front of the Pioneer11 silicon detector which
       had not been taken into account. The original
       exclusion plot for Pioneer11 had therefore excluded for SIMPs
       of arbitrarily large cross-sections.
\item The SKYLAB exclusion region was mis-plotted to exclude upwards from
         the SKYLAB upper diagonal, where\-as it should
         have excluded downwards from the diagonal line, and cut off at
         $\sigma_{Dp}=10^{-22} \mathrm{ cm}^2$ (as shown in Fig. 1).
\end{enumerate}

This new SIMP/CDM window seems rather important since
it overlaps the range of self-scattering cross-sections
($\sigma_{DD}$ in Eq.~(1)) required to solve the
small scale structure problem for WIMPs.    
For some SIMP candidates, $\sigma_{DD}$ and $\sigma_{Dp}$
may be comparable, although this is not required generally.
In these cases, the self-interacting dark matter particles
may be detected directly with suitably-improved space-borne
particle detectors.   

As first reported in (Wandelt {\it et al.}, 2000),
we have revisited the Starkman SIMP dark matter 
analysis, by reinterpreting the Pioneer11 experimental data for the SIMP
hypothesis, plotting the SKYLAB data correctly, and
adding several new exclusion zones to this figure. 
The new exclusion zones are from:
\begin{enumerate}
\item an X-ray quantum calorimeter (XQC)  aboard a sounding
rocket (McCammon {\it et al.}, 1996),
\item the IMAX SIMP search experiment using 
plastic scintillation detectors aboard an
antimatter-search balloon-payload (McGuire {\it et al.}, 1994),
\item the IMP7 and IMP8 satellite-borne silicon
detectors (Me\-waldt {\it et al.}, 2001).
\end{enumerate}

In this work, we have carefully checked the analysis
of the Pioneer11 and the IMP7/8 detectors for the
SIMP-as-DM hypotheses, so we have revised the upper
diagonal of the exclusion zones for these two experiments, and
in Figure 1,  we summarize
for the cosmic ray community the complete SIMP parameter space.

\section{Can Dark Matter Be Observed Through the Crack?}

The area in grey in Figure 1 is the range for the self-scattering
cross-section ($\sigma_{DD}$) required to solve the small
scale structure problem found in WIMP simulations. 
We might first consider SIMPs for which the cross-section with
baryons is similar. This figure shows the comparison
of the presently open or closed parameter space
for the SIMP-baryon scattering cross-section ($\sigma_{Dp}$)
and the SIMP-SIMP scattering cross-section from Equation 1,
with the assumption that the SIMP-baryon scattering cross-section
scales coherently with nucleon number.
The revised exclusion plot for
an alternative assumption of spin-dependent coupling is not reported here.

The basic results are that the XQC detector has
conclusively closed the low-mass area of the 
gray band in the plot.  The reanalysis
of the Pioneer11 and SKYLAB event rates
and the further analysis of the IMP7/8
event rates has reopened the previously-excluded
high-mass area of the gray band in the plot
($M > 10^5$ GeV).

  A possible particle candidate that might exist in the remaining
gray band are Q-balls (coherent states of squarks and
sleptons). Previous Q-ball candidates for dark matter
(see Ku\-sen\-ko and Sha\-po\-shni\-kov, 1998) cannot satisfy the
$\sigma_{DD}/m$ constraints for SIDM (see Eq. 1, above).
Ku\-sen\-ko and Steinhardt (2001) are proposing a variant 
which can serve as SIDM. This Q-ball variant can have
strong self-interacting cross-sections
(corresponding to the gray area) but a very
wide range of interactions with protons, spanning most of the
unexplored regions (plus some range already excluded by existing
measurements). These proposed Q-balls have several unique properties
that make them ideally suited for SIDM.

%\section{Balancing}

%The columns of the last page can be balanced either by using the
%command \verb/\balance/ somewhere in the first column of the last page
%or by explicitely put \verb/\vadjust{\newpage}/ at the correct place.
% without the \verb/ / command

\begin{acknowledgements}
P. McGuire would like to thank Theo\-dore Bo\-wen and the Department
of Physics at the University of Arizona, where this work was initiated.
The IMAX experiment was largely supported by NASA(Goddard Space Flight Center).
Some of this work was also completed during McGuire's residence
at Steward Observatory at the University of Arizona and during
his residence at the Center for Interdisciplinary Studies at
the University of Bielefeld. We thank Dick Mewaldt, Allan Labrador,
Clifford Lopate, and Bruce McKibben for help in understanding
the Pioneer11 detector and the IMP7/8 detectors.  Dan McCammon
and Lyman Page provided us with invaluable inspiration and 
assistance, which led to  the inclusion
of the results of the XQC detector in this analysis.
P. Steinhardt acknowledges the support of 
US Department of Energy grant DE-FG02-91ER40671. 
\end{acknowledgements}

\end{document}